\documentclass[12pt]{article}%
\usepackage{amsfonts}
\usepackage{physics}
\usepackage{bm}
\usepackage{bbm}
\usepackage{subfiles}
\usepackage{amsmath}
\usepackage{graphicx}
\usepackage{hyperref}
\usepackage{verbatim}
\usepackage{color}
\usepackage{tikz}
\usepackage{pgfplots}
\usepgfplotslibrary{fillbetween}
\usetikzlibrary{patterns}
\usepackage{enumerate}
\usepackage{amsfonts}
\usepackage{amssymb}
\usepackage{authblk}
\usepackage[hang,flushmargin]{footmisc}
\usepackage{lipsum}
\usepackage[font=footnotesize]{caption}
\usepackage{subcaption}
\usepackage[utf8]{} 
\usepackage{empheq}

\csname @addtoreset\endcsname{equation}{section}
 \newcommand{\badat}{\begin{alignedat}}
 \newcommand{\eadat}{\end{alignedat}}

\usepackage{csquotes}
\usepackage[backend=biber, hyperref,backref,backrefstyle=none,style=numeric-comp, sorting=none,maxbibnames=5,url=false]{biblatex}

\DefineBibliographyStrings{english}{%
	backrefpage = {cited on page},%
	backrefpages = {cited on pages}%
}
\DeclareFieldFormat{doi}{%
	\hfil\penalty90\hfilneg\space DOI\addcolon\addnbspace
	\ifhyperref
	{\href{https://doi.org/#1}{\nolinkurl{#1}}}
	{\nolinkurl{#1}}}

\def\be{\begin{eqnarray}}
\def\ee{\end{eqnarray}}
\def\beann{\begin{eqnarray*}}
\def\eeann{\end{eqnarray*}}
\def\beq{\begin{equation}}
\def\eeq{\end{equation}}
\def\ba{\begin{array}}
\def\ea{\end{array}}
\def\ben{\begin{enumerate}}
\def\een{\end{enUmerate}}
\def\bea{\begin{eqnarray}}
\def\eea{\end{eqnarray}}

\def\5{\bar }
\def\6{\partial }
\def\7{\hat }
\def\4{\tilde }

\def\cD{\mathcal{D}}
\def\cE{\mathcal{E}}

\def\cH{\mathcal{H}}

\def\cL{\mathcal{L}}

\def\cR{\mathcal{R}}

\newcommand{\LL}{\mathcal{L}}
\newcommand{\mum}{{\mu^-}}
\newcommand{\mup}{{\mu^+}}

\newcommand{\LLp}{{\mathcal{L}}^+}
\newcommand{\LLm}{{\mathcal{L}}^-}
\newcommand{\euclidean}[1]{#1_\mathrm{E}}

\newcommand{\E}{\mathrm{E}}
\newcommand{\etapm}{\eta^\pm}

\newcommand{\mean}[1]{\left\langle #1 \right\rangle} 

\newcommand{\LLpm}{{\mathcal{L}^\pm}}
\newcommand{\HHpm}{H^\pm}

\renewcommand{\d}{\partial}

\renewcommand{\tilde}{\widetilde}
\renewcommand{\hat}{\widehat}

\begin{document}

\title{\vspace{-70pt} \Large{\sc 1/c deformations of AdS$_3$ boundary conditions and the Dym hierarchy}\vspace{10pt}}
\author[a,b,c]{\normalsize{Kristiansen Lara}
\footnote{\href{mailto:klara@cecs.cl}{klara@cecs.cl}}}
\author[c,d]{\normalsize{Miguel Pino}\footnote{\href{mailto:miguel.pino.r@usach.cl}{miguel.pino.r@usach.cl}}}
\author[c]{\normalsize{Francisco Reyes}\footnote{\href{mailto:francisco.reyes.z@usach.cl}{francisco.reyes.z@usach.cl}}}

\affil[a]{\footnotesize\textit{Centro de Estudios Científicos (CECs), Av. Arturo Prat 514, Valdivia, Chile.}}
\affil[b]{\footnotesize\textit{Facultad de Ingeniería, Arquitectura y Diseño, Universidad San Sebastián, sede Valdivia, General Lagos 1163, Valdivia 5110693, Chile.}}
\affil[c]{\footnotesize\textit{Departamento de F\'isica, Universidad de Santiago de Chile, Av. V\'ictor Jara 3493, Santiago, Chile.}}
\affil[d]{\footnotesize\textit{Center for Interdisciplinary Research in Astrophysics and Space Exploration (CIRAS),
Universidad de Santiago de Chile, Av. Libertador Bernardo O’Higgins 3363, Santiago, Chile.}}

\date{}

\maketitle
\thispagestyle{empty}
\begin{abstract}
  This work introduces a novel family of boundary conditions for AdS$_3$ General Relativity, constructed through a polynomial expansion in negative integer powers of the Brown-Henneaux central charge. The associated dynamics is governed by the Dym hierarchy of integrable equations. It is shown that the infinite set of Dym conserved charges generates an abelian asymptotic symmetry group. Additionally, these boundary conditions encompass black hole solutions, whose thermodynamic properties are examined.

\end{abstract}

\newpage

\begin{small}
{\addtolength{\parskip}{-2pt}
 \tableofcontents}
\end{small}
\thispagestyle{empty}
\newpage

\section{Introduction}

The holographic principle \cite{Thorn:1991fv,tHooft:1993dmi,Susskind:1994vu} is a valuable tool for exploring General Relativity and gauge theories. For any realization of holography, it is crucial to define well-posed boundary conditions for the dynamical fields, since the corresponding asymptotic symmetries match the global symmetries of the boundary theory. The precursory example by Brown and Henneaux in three-dimensional General Relativity with negative cosmological constant \cite{Brown:1986nw}, highlights the relevance of boundary conditions and the identification of asymptotic symmetries. Although the best known realization of holography, the  AdS/CFT correspondence \cite{Maldacena:1997re,Witten:1998qj,Gubser:1998bc} was proposed in a string theory framework, the Brown-Henneaux example provides valuable insights into holography without relying on string theory.

The consequences of choosing a wide variety of boundary conditions for the gravitational field on several different scenarios has been extensively explored in the literature. The seminal example concerns four-dimensional asymptotically flat spacetimes, for which the infinite dimensional BMS$_4$ group~\cite{Bondi:1962px,Sachs:1962Asymptotic} arises as the asymptotic symmetries at null infinity. This classic result has been scrutinized and extended for several decades~\cite{Penrose:1965am,Geroch1977,Barnich:2010eb,Barnich:2010ojg,Barnich:2011mi,Compere:2011ve,Barnich:2012xq,Bagchi:2012xr,Barnich:2012aw,Barnich:2013sxa,Campiglia:2014yka,Barnich:2014kra,Duval:2014uva,Penna:2015gza,Fuentealba:2015wza,Barnich:2015jua,Barnich:2015uva,Gomis:2015ata,Batlle:2017llu,Grumiller:2017sjh,Troessaert:2017jcm,Barnich:2017ubf,Batlle:2017yuz,Hijano:2018nhq,Henneaux:2018cst,Bunster:2018yjr,Compere:2019bua,Compere:2020lrt,Fuentealba:2020zkf,Fuentealba:2020aax,Campiglia:2020qvc,Fuentealba:2021yvo,Fuentealba:2022yqt,Fuentealba:2022xsz,Barnich:2022bni,Bekaert:2022ipg,Fuentealba:2023syb}.

The choice of boundary conditions have been used to explore dualities in a broader context. A nonexhaustive list includes the Kerr/CFT duality~\cite{Guica:2008mu} (for a review, see \cite{Compere:2012jk}), Schrödinger-invariant spacetimes~(see \cite{Son:2008ye} and references thereof), dualities with condensed matter physics~(for a review, see \cite{Hartnoll:2016apf}), among many others. It also plays a role in the context of the soft hair proposal~\cite{Hawking:2016,Hawking:2016sgy}: it amounts to choose the behavior of the gravitational field in the near-horizon region of rotating black holes~\cite{Donnay:2015abr,Donnay:2016ejv,Giribet:2021zie,Giribet:2023xom}. The same ideas has been applied to cosmological horizons~\cite{Kehagias:2016zry,Bonga:2020fhx,Enriquez-Rojo:2020miw,Enriquez-Rojo:2021blc} and plays a key role in the recent celestial holography program (for reviews, see e.g.,~\cite{Prema:2021sjp,Pasterski:2023ikd,Donnay:2023mrd} and references therein.)

Three-dimensional gravity is a successful toy model to explore the holographic consequences of a variety of boundary conditions. Because it is devoid of bulk degrees of freedom, the complete dynamical content is captured by boundary excitations. Out of simplicity, throughout the paper we will be focusing on pure AdS$_3$ General Relativity. The aforementioned Brown-Henneaux boundary conditions is the precursor example, on which the asymptotic symmetry algebra corresponds to two copies of Virasoro algebra with central charge
\begin{equation}
  c_\mathrm{BH}=\frac{3 \ell}{2G},
\end{equation}
where $\ell$ stands for the AdS radius, while $G$ represents Newton's constant. However, many others possibilities for asymptotic behavior have been explored in the literature~\cite{Henneaux:2002wm,Liu:2009kc,Henneaux:2011hv,Henneaux:2010fy,Compere:2013bya,Troessaert:2013fma,Avery:2013dja,Donnay:2015abr,Afshar:2015wjm,Grumiller:2016pqb,Valcarcel:2018kwd,Henneaux:2019sjx,Alessio:2020ioh,Ozer:2019nkv,Ozer:2021wtx}.

Following \cite{Bunster:2014mua,Perez:2016vqo}, this paper concerns the possibility of exploiting the near boundary behavior of the lapse and shift functions, while leaving the spatial part of the metric in standard AdS$_3$ asymptotic form. To make the following discussion more precise, consider a manifold with coordinates $(t,\rho,\phi)$, where $0\leq t<\infty$, $0\leq\phi\leq 2\pi$ and $0\leq \rho <\infty$. The boundary is located at $\rho\to\infty$. The formulation of gravitational boundary conditions can be most simply achieved within the Chern-Simons formulation of AdS$_3$ General Relativity~\cite{Achucarro:1986uwr,Witten:1988hc} (See \ref{Appendix: Section: Chern-Simons Gravity} for notation and conventions). The asymptotic structure of the gravitational field is encoded in the behavior of the two $sl(2,\mathbb{R})$ connections $A^\pm$ near the boundary. As explained in \cite{Coussaert:1995zp}, asymptotically AdS spacetimes are described by connections of the form $A^\pm=b^{-1}_\pm a^\pm b_\pm+b^{-1}_\pm db_{\pm}$, where the group element $b_{\pm}(\rho)$ gauges away the radial dependence. The spatial part of the auxiliary connection is given by
\begin{subequations}
	\label{bcspatial}
\begin{align}
	\label{eq: aphi}
  a_\phi^\pm&=L_{\pm 1}-\frac{12\pi}{c_{\mathrm{BH}}}\cL^\pm L_{\mp 1}\,,\\
  a_{\rho}^\pm&=0\,,
\end{align}
\end{subequations}
where $\cL^\pm$ are functions of $t$ and $\phi$.

As shown in \cite{Bunster:2014mua}, the most general gauge transformation preserving the form of \eqref{bcspatial} has a gauge parameter of the form  
\begin{equation}
	\label{eq: Lambda}
  \Lambda^\pm[\eta^\pm]=\pm \left(\eta^\pm a_\phi^\pm   \mp \d_\phi \eta^\pm L_0+ \frac{1}{2} \d_\phi^2 \eta^\pm L_{\mp 1}   \right)\,.
\end{equation}
Regarding the temporal component $a_t^\pm$, its choice should ensure that the spatial boundary conditions \eqref{bcspatial} are preserved under temporal evolution. Therefore, it should also follow the form given by equation \eqref{eq: Lambda}, 
\begin{equation}
\label{at}
a^\pm_t=\frac{1}{\ell}\Lambda^\pm[\mu^\pm]\,.
\end{equation}
According to the map among Chern-Simons and metric fields, the functions $\mu^\pm$ encode the asymptotic behavior of the lapse and shift functions.

It is important to highlight that $\mu^\pm$ should be chosen consistently in order to describe well defined boundary conditions for the gravitational field. For example, Brown-Henneaux boundary conditions \cite{Brown:1986nw} are obtained with the choice $\mu^\pm=1$, yielding $a_t^\pm=\pm\frac{1}{\ell}a_\phi^\pm$. In this case, the gauge parameters $\eta^\pm$ fulfill a chiral boson equation, rendering the asymptotic symmetry group infinite-dimensional and spanning two copies of Virasoro algebra with central charge $c_\mathrm{BH}$. Another noteworthy example was given in \cite{Perez:2016vqo}, where $\mu^\pm$ are chosen as functions of $\cL^\pm$ and its derivatives; by fixing $\mu^\pm$ as the $n$-th Gelfand-Dikii polynomial~\cite{Gelfand:1975rn} of $\cL^\pm$, Einstein's equation reduces to the $n$-th member of the Korteweg-de Vries (KdV) hierarchy of integrable equations, unveiling a connection between integrable systems and 3D gravity. The integrability of the KdV hierarchy translates into an infinite dimensional and abelian asymptotic symmetry algebra generated by the KdV conserved charges. This connection with KdV hierarchy has been further explored~\cite{Fuentealba:2017omf,Gonzalez:2018jgp,Melnikov:2018fhb,Ojeda:2019xih,Erices:2019onl,Ojeda:2020bgz,Dymarsky:2020tjh}. The relation between 3D gravity and integrable systems was extended to include the Ablowitz-Kaup-Newell-Segur (AKNS) system~\cite{Cardenas:2021vwo}.

As shown by the previous examples, there is further ground for exploration when choosing state-dependent Lagrange multipliers and their holographic implications. However, such a choice is not arbitrary since it must meet the necessary requirements of well-defined boundary conditions. Integrable systems provide the necessary tools to explore the landscape of possibilities and might serve as a guideline to explore nonstandard holography. In this sense, as a first step toward this objective, this paper aims to enlarge the aforementioned landscape of boundary conditions for AdS$_3$ gravity by using recursive procedures inspired by the construction of the AKNS system.

We will construct a new class of functions $\mu^\pm$, dependent on $\cL^\pm$ and its derivatives, that yield an infinite family of well-defined boundary conditions labeled by a positive integer $N$. The corresponding dynamical equations are those of the Harry Dym hierarchy of nonlinear integrable equations~\cite{Kruskal1975}. The asymptotic symmetries are generated by an infinite tower of conserved charges, forming an abelian algebra under the Poisson bracket, as expected from the integrability properties of the dynamical equations.

Inspired by the ansatz of AKNS~\cite{Ablowitz:1974ry}, the key step in the construction is the assumption that the functions $\mu^\pm$ can be expanded as a polynomial in negative powers of the Brown-Henneaux central charge $c_\mathrm{BH}$. As explained below, the coefficients of such polynomials are constructed following a recursion relation, much in the spirit of integrable systems (see e.g., \cite{Drazin:1989qi}).

The paper is structured as follows: Section \ref{construction} details the construction of boundary conditions, while Section \ref{consistency} shows their well-defined nature. Some final comments and appendixes are found at the end of the paper.

\section{Constructing the family of boundary conditions} \label{construction}

The aim of this section is to construct an infinite family of possible choices for the functions $\mu^\pm$ in \eqref{at} as local functions of $\cL^\pm$ and its derivatives. The starting point is the spatial behavior of the connection \eqref{bcspatial}, supplemented with the temporal component given by \eqref{at}. The first-order Einstein equations translates in the Chern-Simons formulation as two independent copies of the zero-curvature condition $\d_t a^\pm_\phi-\d_\phi a^\pm_t+\left[ a^\pm_t ,a^\pm_\phi \right]~=~0$, yielding the evolution equations 
\begin{equation}
  \pm\ell \d_t \cL^\pm = 2 \cL^\pm \d_\phi \mu^\pm  +  \d_\phi \cL^\pm\mu^\pm - \frac{c_{\mathrm{BH}}}{24\pi} \d^3_\phi \mu^\pm     \,.
\end{equation}
These dynamical equations will serve as a guide for constructing the sought functions $\mu^\pm$. In this regard, the above is conveniently expressed as follows
\begin{equation}\label{Ldot}
  \pm \ell\d_t \cL^\pm = \left( \cD^\pm  - c \cE    \right)\mu^\pm,
\end{equation}
where the operators $\cD^\pm$ and $\cE$ are defined as
\begin{equation}
	\label{eq: Hamiltonian operators}
  \cD^\pm \equiv  2 \cL^\pm \d_\phi + \d_\phi \cL^\pm,\quad\quad\quad \cE \equiv \d_\phi^3.
\end{equation}
These two differential operators are Hamiltonian, in the sense that they are linear, anti-symmetric and fulfill Jacobi identity~\cite{olver2000applications}. As shown below, these form the bi-Hamiltonian structure of the Harry Dym hierarchy. The central charge in \eqref{Ldot} is rescaled $c \equiv c_{\mathrm{BH}}/24\pi$ to avoid $24 \pi$ terms in the forthcoming expressions.

As mentioned in the introduction, the main assumption is that the functions $\mu^\pm$ can be expanded in negative powers of the central charge $c$ up to some fixed integer power $N\ge 1$
\begin{equation}
	\label{expansion}
  \mu^\pm=\mu^\pm_0+\sum^{N}_{n=1}\mu^\pm_n \frac{1}{c^{N-n+1}}\,.
\end{equation}
The positive integer $N$ is arbitrary and can be chosen freely. It plays the role of labeling the family of possible choices for $\mu^\pm$; different values for $N$ corresponds to inequivalent boundary conditions. Although the integer $N$ can have different values on both sectors $\pm$, here are chosen the same for simplicity. 

When substituted in the equation of motion \eqref{Ldot}, the polynomial ansatz \eqref{expansion} allows for grouping coefficients with the same power of $c$, resulting in relations among the coefficients $\mu^\pm_n$. Firstly, the terms linear in $c$ renders the condition
\begin{equation}
	\label{emu0}
  \cE \mu_{0}^\pm=0\,,
\end{equation}
which fixes the first coefficient $\mu_0^\pm$. The solution to \eqref{emu0}, respecting periodic boundary conditions, is given by a constant $\mu_0^\pm=\alpha_0^\pm$.

The $c^0$ terms, involving the time derivative of $\cL^\pm$, yield the evolution equation
\begin{equation}
\label{eq: Dym hierarchy}
  \pm \ell\d_t\cL^\pm =\left(  \cD^\pm \mu_0^\pm -  \cE \mu_N^\pm \right)\,.
\end{equation}
Since $\mu^\pm_0=\alpha_0^\pm$, the first term on the right-hand side of the equation becomes $\alpha_0^\pm \partial_\phi \cL^\pm$. Therefore, the dynamical equation corresponds to the chiral boson deformed by the second term, which depends on $\mu_N^\pm$. To determine the form of $\mu_N^\pm$, we can use the remaining $c^{-1}$ to $c^{-N}$ terms, as they imply the following relations
\begin{subequations}
	\label{eq: Conditions mu}
\begin{align}
  \cD^\pm \mu^\pm_1&=0\,, \\
  \label{eq: Bi-Hamiltonian relationship}
  \cD^\pm \mu^\pm_{n+1} &= \cE  \mu^\pm_{n}\,,\quad n\geq 1\,.
\end{align}
\end{subequations}
The former differential equation governs $\mu_1^\pm$, while the latter establishes a recursive relation for deriving $\mu^\pm_{n+1}$ from its predecessor. These relations allows to iteratively construct all coefficients $\mu^\pm_n$ as functions of $\LLpm$ and its spatial derivatives. Once the coefficients $\mu_n^\pm$ are determined, the function $\mu^\pm$ is fully specified by the summation in equation \eqref{expansion}, thereby completely defining the boundary conditions.

It is important to remark that solving the recursion relation leads to integration constants. These arise due to the linearity of the differential equation \eqref{eq: Bi-Hamiltonian relationship}; the solution for $\mu_{n+1}^\pm$ consists of a homogeneous component with an integration constant, as well as a particular solution that depends on the previous coefficient $\mu_n^\pm$. To shed light on this observation, lets examine the role of the differential operators in the recursion relation. Consider the inverse of the operator $\cD^\pm$, given by
\begin{align}
	{\cD^\pm}^{-1} [X(\phi)] = \frac{1}{2 \sqrt {\cL^\pm}}\d_\phi^{-1}\left(\frac{X(\phi)}{\sqrt{ \cL^\pm}} \right)\,.
\end{align}
It allows to define the recursion operator $\cR^\pm\equiv {\cD^\pm}^{-1} \circ \cE $, yielding 
\begin{align}
	\cR^\pm [X(\phi)] = \frac{1}{2 \sqrt {\cL^\pm}}\partial_\phi^{-1}\left(\frac{1}{\sqrt{\LLpm}}\partial_\phi^3 X(\phi)\right)\,.
\end{align}
Accordingly, the recursion relation \eqref{eq: Bi-Hamiltonian relationship} can then be expressed as 
\begin{align}\label{eq: mun+1}
	\mu_{n+1}^\pm=\cR^\pm [\mu_n^\pm]+\cR^\pm [0]\,.
\end{align}
The first term on the right-hand side produces the particular solution, whereas the subsequent term corresponds to the homogeneous one, i.e.,  $\cR^\pm [0]=\frac{\alpha}{2 \sqrt{\cL^\pm}}$, where $\alpha$ is an integration constant. Then, the complete solution to the differential equation \eqref{eq: Bi-Hamiltonian relationship} is given by
\begin{align}
	\mu_{n+1}^\pm=\cR^\pm [\mu_n^\pm]+\frac{\alpha^\pm_{n+1}}{2 \sqrt{\cL^\pm}}\,.
\end{align}
This result can be successively used to derive an explicit expression for any of the coefficients $\mu_{n}^\pm$,
\begin{align}
	\mu_{n}^\pm=\sum_{i=1}^{n}\alpha_{i}^\pm {\cR^\pm}^{n-i}\left[\frac{1}{2 \sqrt{\cL^\pm}}\right]     \,,\quad n\geq 1\,.
\end{align}

At this point, it is important to highlight a remarkable property of the recursion operator $\cR^\pm$; the successive action of $\cR^\pm$ on the expression $\frac{1}{2 \sqrt{\cL^\pm}}$ consistently yields functional derivatives of quantities $H_n^\pm$, according to
\begin{align}\label{magia}
	{\cR^\pm}^{n}\left[\frac{1}{2 \sqrt{\cL^\pm}}\right]=\frac{\delta H_{n+1}^\pm}{\delta \cL}     \,.
\end{align}
The reason behind this lies in the properties of the operators $\cD^\pm$ and $\cE$; these are not just Hamiltonian operators but are \emph{compatible} among each other, meaning that any linear combination of them also renders a Hamiltonian operator. This compatibility ensures \eqref{magia}, as demonstrated in \cite{olver2000applications}. Such a property is common in bi-Hamiltonian integrable systems, proving useful as it allows the construction of an infinite tower of conserved quantities of the system. As illustrated below, the quantities $H_n^\pm$ found in our construction are conserved charges and serve as generators for the asymptotic symmetry algebra. An explicit list of the first $H_n^\pm$ along with its functional derivatives can be found in Appendix \ref{Appendix: Subsection: List}.

According to the previous discussion, the coefficients $\mu_n^\pm$ take the form
\begin{align}
	\label{eq: Chemical potential summation}
	\mu_n^\pm=\sum_{i=1}^{n}\alpha_{n-i+1}^\pm\fdv{H_i^\pm}{\LLpm}\,,\quad n\geq 1\,,
\end{align} 
rendering the functions $\mu^\pm$ as 
\begin{align}
	\label{eq: Polynomial ansatz mu general}
	\mu^\pm=\alpha_0^\pm+\sum_{n=1}^{N}\sum_{i=1}^{n}\frac{1}{c^{N-n+1}}\alpha_{n-i+1}^\pm\fdv{H_i^\pm}{\LLpm}\,.
\end{align}

While the above expression is sufficient to specify the boundary conditions, for the sake of simplicity, we aim to make specific choices for the arbitrary constants $\alpha_n^\pm$. As mentioned earlier, opting for $\alpha^\pm_{i}=\delta_{i,0},\, i\ge 0,$ yields the Brown-Henneaux boundary conditions, since all the $1/c$ terms in the $\mu^\pm$ functions disappear. Another viable option is $\alpha_{0}^\pm=0$ and $\alpha_{i}^\pm=\delta_{i,1},\, i\ge 1$. This selection has the virtue of extending the asymptotic symmetries to include Lifshitz scaling. However, this possibility will be explored in a separate study.

For the current analysis, we set $\alpha_{0}^\pm=1$ and $\alpha_{i}^\pm=\delta_{i,1},\, i\ge 1$. With these choices, the coefficients $\mu_n^\pm$ take on the following form
\begin{align}
	\label{eq: mu as a variation}
	\mu^\pm_n=\fdv{\HHpm_n}{\LLpm}\,.
\end{align}
(Note that within this choice, the same expression also holds for $\mu_0^\pm$ due to the relation $\mu_0^\pm=\frac{\delta H_0^\pm}{\delta \cL^\pm}$, where $H_0^\pm \equiv \int \cL^\pm\, d\phi$). The functions $\mu^\pm$ result in
\begin{align}
	\label{eq: mu functional derivative}
	\mu^\pm=1+\sum_{n=1}^{N}\frac{1}{c^{N-n+1}}\fdv{\HHpm_n}{\LLpm}\,.
\end{align}
The latter expression reinforces the title of this work: these boundary conditions can be interpreted as the standard Brown-Henneaux case, denoted by the ``1" in the latter expression, deformed by suitable terms in negative powers of the central charge, symbolized by the sum. It is worth highlighting that this construction is valid for any value of the central charge, and hence, it is not a perturbative result.

To conclude this section, let us briefly delve into the dynamical equation \eqref{eq: Dym hierarchy}. Given the choices outlined above for the arbitrary constants $\alpha_n^\pm$, the simplest scenario arises for $N=1$, reducing \eqref{eq: Dym hierarchy} to the well-known Dym's equation
\begin{align}
\label{eq: Dym equation}
\pm\ell\partial_t\LLpm&=\partial_\phi\LLpm-\partial_\phi^3\left(\frac{1}{2\sqrt{\LLpm}}\right).
\end{align}
This non-linear integrable evolution equation was initially discovered in an unpublished manuscript by Harry Dym and later revisited in \cite{Kruskal1975}. It manifests in various physical scenarios, including the classical string~\cite{Sabatier:1979yg,Yi-Shen1982} and problems in fluid mechanics, specifically governing the dynamics at the interface between two fluids with different viscosities, known as the Saffman-Taylor problem~\cite{Saffman-Taylor}. Dym's equation admits a diverse set of traveling solitons~\cite{Fuchssteiner_1992,Mokhtari2011,SolitonsHD, Novikov1999,RoyChowdhuryMukherjee1984,Bordag:1995dp,alezGaxiola2018ITERATIVEMF,XIAO2019123248,Assabaai2022ExactSO,LI2020106276,Li2017,DMITRIEVA199365,shunmugarajan2016efficient}. An extension to higher dimensions was later uncovered in \cite{KONOPELCHENKO198415}. The resulting higher-order differential equations obtained for $N>1$, collectively referred to as the Dym hierarchy, are given by
\begin{equation}\label{dymhierarchy}
 \pm\ell\partial_t\LLpm =\partial_\phi\LLpm-\partial_\phi^3\left(\frac{\delta H_N^\pm  }{\delta \cL^\pm  }\right).
\end{equation}

Concerning the integrability of Dym's hierarchy, it is important to highlight that each equation within the hierarchy shares an identical infinite tower of conserved quantities $H_n^\pm$~\cite{Magri1978,Infinite_Number}. To demonstrate this, consider equations \eqref{eq: Bi-Hamiltonian relationship} and \eqref{eq: Chemical potential summation}, which give rise to the recursion relation
\begin{align}
\label{eq: Hierarchy in bi-Hamiltonian form}
\cD^\pm\left(\fdv{\HHpm_{n+1}}{\LLpm}\right)=\cE\left(\fdv{\HHpm_n}{\LLpm}\right),\quad n\geq 0.
\end{align}
By virtue of the latter, we can express the $N$-th member of the hierarchy \eqref{eq: Dym hierarchy} in two distinct Hamiltonian forms
\begin{align}
\label{eq: Recurrence relation}
\pm\ell\partial_t\LLpm=\cD^\pm\left(\fdv{H_0^\pm}{\LLpm}-\fdv{\HHpm_{N+1}}{\LLpm}\right)=\cE\left(\fdv{H_{-1}^\pm}{\LLpm}-\fdv{H_N^\pm}{\LLpm}\right),
\end{align}
(where $\HHpm_{-1}=-\frac{1}{2}\int d\phi\, \left(\partial_\phi^{-1}\LLpm\right)^2$), thereby establishing each element of Dym's hierarchy as a bi-Hamiltonian system. Furthermore, the differential operators $\cD^\pm$ and $\cE$ enable the construction of two nonequivalent Poisson structures
\begin{align}
\label{eq: Poisson brackets}
\pb{F}{G}_{\cD}=\int d\phi\,\fdv{F}{\LLpm}\cD^\pm\left(\fdv{G}{\LLpm}\right),\quad \pb{F}{G}_{\cE}=\int d\phi\,\fdv{F}{\LLpm}\cE\left(\fdv{G}{\LLpm}\right).
\end{align}
A well-established result follows from the recursion relation \eqref{eq: Hierarchy in bi-Hamiltonian form}, asserting that all quantities $H_n^\pm$ Poisson commute with each other, utilizing either of the two defined Poisson brackets (see \cite{olver2000applications}),
\begin{align}\label{involution}
\pb{H_n^\pm}{H_m^\pm}_{\cD,\cE}=0, \quad \forall\, n,m\ge 0.
\end{align}
The involution of the quantities $H_n^\pm$ demonstrates that each of them serves as a conserved charge for the evolution equation, conclusively establishing the integrability of each equation within the hierarchy.

\section{Consistency of the boundary conditions}\label{consistency}

Well-defined boundary conditions must meet three criteria~\cite{Henneaux:1985tv}. First, they must yield a differentiable action principle. Second, they should permit an asymptotic symmetry group with finite generators. Lastly, it is desirable for them to encompass appealing gravitational configurations, such as black holes.

The objective of this section is to demonstrate that the chosen boundary conditions \eqref{bcspatial}\eqref{at} with \eqref{eq: mu functional derivative} fulfill all three aforementioned criteria.

\subsection{Action principle}\label{Subsection: Integration of the boundary term}

The Hamiltonian Chern-Simons action is given by
\begin{align}
	I_H^\pm=-\frac{k}{4\pi}\int\, dt\,d^2x\,\epsilon^{ij}\mean{{A}_i^\pm \dot{A}_j^\pm-A_t^\pm F_{ij}^\pm}+B^\pm\,,
\end{align}
where the boundary term $B^\pm$ fulfills 
\begin{align}
  \delta B^\pm=-\frac{k}{2\pi}\int_{\rho\to\infty}dt\,d\phi\, \mean{A_t^\pm \delta A_\phi^\pm}\,,
\end{align}
By using the asymptotic behavior of the fields \eqref{bcspatial} and \eqref{at}, the variation of this surface integral reads
\begin{align}
	\delta B^\pm=\mp\frac{k}{4\pi c\ell}\int_{\rho\to\infty}dt\,d\phi\,\mu^\pm\delta\cL^\pm\,.
\end{align}
By means of equation \eqref{eq: mu functional derivative}, the boundary term becomes integrable, rendering
\begin{align}
	B^\pm_N=\mp\int \frac{dt}{\ell}\left(\HHpm_0+\sum_{n=1}^N\frac{1}{c^{N-n+1}}\HHpm_n\right)\,.
\end{align}
The subscript $N$ is included as a reminder that the boundary term is different for distinct choices of $N$.

Since the full gravitational action is the difference of two Chern-Simons functionals $I=I_H^+-I_H^-$, then the full boundary term $B_N=B_N^+-B_N^-$ becomes 
\begin{align}
	\label{eq: Boundary term CS}
	B_N=-\int \frac{dt}{\ell}\left[H_0^++H_0^-+\sum_{n=1}^{N}\frac{1}{c^{N-n+1}}\left(H_n^++H_n^-\right)\right]\,.
\end{align}

Therefore, the action principle is differentiable.

\subsection{Asymptotic symmetries}

Here, we show that the previously identified conserved quantities $H_n^\pm$ serve as canonical generators for the asymptotic symmetry group associated with the given boundary conditions.

To begin, let us examine the asymptotic behavior of the angular component of the connection $a_\phi$, shown in \eqref{eq: aphi}. As mentioned in the introduction, the most general gauge transformation that preserves such behavior has a parameter of the form \eqref{eq: Lambda}~\cite{Bunster:2014mua}. The corresponding transformation of the dynamical fields $\cL^\pm$ can be expressed as 
\begin{align}
	\label{eq: Infinitesimal transformation L}
	\pm\delta \cL^\pm=\left(\cD^\pm-c\cE\right)\eta^\pm\,.
\end{align}
At this point, the typical approach involves exploring the preservation of the temporal component of the connection $a_t^\pm$, often leading to additional conditions on the functions $\eta^\pm$. However, we opt for an alternative approach. First, by following the methodology outlined in the previous section, we construct a family of possible choices for $\eta^\pm$. Subsequently, we determine the canonical charges and their algebra. Towards the end of the section, we demonstrate that these transformations also respect the form of $a_t^\pm$.

The construction employs the same methodology as in section \ref{construction}, where $\eta^\pm$ is treated as an expansion in negative powers of the central charge up to some arbitrary integer $M$. It is then inserted into \eqref{eq: Infinitesimal transformation L}, and the coefficients of the expansion are determined by solving order by order. Following the selection of arbitrary constants in a similar fashion to the previous section, this construction yields the following result
\begin{align}
	\label{eq: Eta expansion}
	\eta^\pm=\epsilon^\pm_0+\epsilon^\pm_M\sum_{m=1}^M\frac{1}{c^{M-m+1}}\fdv{\HHpm_m}{\LLpm}\,,
\end{align}
where $\epsilon_0$ and $\epsilon_M$ are arbitrary infinitesimal constants.

From a canonical perspective, the variation of the generators for these transformations can be expressed as
\begin{align} \label{eq: Delta Q}
  \delta Q^\pm[\eta^\pm]=\mp \frac{1}{\ell} \int d\phi\,\etapm\delta \LLpm\,.
\end{align}
This expression is integrable due to \eqref{eq: Eta expansion}. Consequently, we designate the generators as $Q_0^\pm\equiv Q^\pm[\epsilon^\pm_0=1,\epsilon^\pm_M=0]$ and $Q_M^\pm\equiv Q^\pm[\epsilon^\pm_0=0,\epsilon^\pm_M=1]$, leading to the following identifications
\begin{subequations}\label{las cargas}
\begin{align}	
  Q_0^\pm &=\mp \frac{1}{\ell}H_0^\pm\, ,\\
  Q_M^\pm&=\mp \frac{1}{\ell} \sum_{m=1}^M\frac{1}{c^{M-m+1}}\HHpm_m\,.
\end{align}
\end{subequations}
It is important to emphasize that $M$ can take any positive integer value. Therefore, the generator $Q_M^\pm$ corresponds to an infinite family of generators, each labeled by a distinct value of $M$. According to \eqref{eq: Infinitesimal transformation L}, the generators induce the following transformations for the dynamical fields
\begin{subequations}\label{transformations}
  \begin{align}
    	\pm\delta_0\cL^\pm&=\partial_\phi\cL^\pm\, ,\\
  	\pm\delta_M\cL^\pm&=-\cE\left(\frac{\delta H_M^\pm}{\delta \cL^\pm}\right)\,.
\end{align}
\end{subequations}
With the generators now identified, we can proceed to establish their algebra. For this purpose, the following relation proves to be useful~\cite{Brown:1986nw}
\begin{align}
	\label{eq: ASA}
	\left\{Q^\pm\left[\eta_1^\pm\right],Q^\pm\left[\eta_2^\pm\right]\right\}=\delta_1 Q^\pm_2 \,.
\end{align}
It is important to distinguish the above Poisson bracket from those defined in \eqref{eq: Poisson brackets}. The former corresponds to the gravitational phase space, while the latter relates to the bi-Hamiltonian structure of the Dym hierarchy. These are not necessarily equivalent. However, through the utilization of relation \eqref{eq: ASA} and considering the transformations in \eqref{transformations}, it is straightforward to show that
\begin{align}
  \left\{Q^\pm_M,Q^\pm_{\bar M}\right\}=\sum_{m=1}^{\bar M} \frac{1}{c^{\bar M -m +1}}\int d\phi\, \frac{\delta H^\pm_m}{\delta \cL^\pm } \cE \left( \frac{\delta H_M^\pm}{\delta \cL^\pm } \right) \,.
\end{align}
The right-hand side of the latter equation corresponds to $\pb{H_m^\pm}{H_M^\pm}_{\cE}$, which vanishes due to \eqref{involution}. Consequently, the generators satisfy an abelian algebra, as expected from the integrability of the dynamical equations. A similar procedure can be executed with $Q_0^\pm$, leading to the following generator algebra
\begin{align}
	\label{eq: Asymptotic algebra}
  \left\{Q_M^\pm,Q_{\bar{M}}^\pm\right\}=0\quad \forall\,M,\bar M \geq 0. 
\end{align}

To finalize this section, we show that the aforementioned asymptotic gauge transformations also preserve the form of the temporal component $a_t^\pm$. The corresponding gauge transformation is given by
\begin{align}
	\label{eq: Gauge transformation t}
	\delta a_t^\pm=\partial_t \Lambda^\pm+\comm{a_t^\pm}{\Lambda^\pm}\,.
\end{align}
As mentioned previously, due to the temporal derivative acting on the gauge parameter, it is expected that the latter equation will further impose constraints on the form of $\eta^\pm$. For instance, in the case of Brown-Henneaux boundary conditions, a consecuence of \eqref{eq: Gauge transformation t} is that the functions obey a chiral boson equation. In the case at hand, it turns out that the equation \eqref{eq: Gauge transformation t} implies no further conditions on the functions $\eta^\pm$ but rather reduces to a combination of the equation of motion \eqref{eq: Dym hierarchy} and the transformation of the field \eqref{transformations}, thereby becoming an identity. To prove this statement, we follow~\cite{Cardenas:2021vwo} to establish a general relation applicable to any gauge transformation of the form \eqref{eq: Lambda} along with \eqref{eq: Eta expansion}.

First, consider two gauge transformations of parameters $\Lambda^\pm$ and $\bar \Lambda^\pm$, whose action on an auxiliary connection $a$ is given by
\begin{align} \label{deltas}
  \delta a =& d \Lambda + \left[ a,\Lambda \right]\, ,\\
  \bar \delta a =& d \bar \Lambda + \left[a,\bar \Lambda \right]\,.
\end{align}
A straightforward calculation reveals that the commutator of two such gauge transformations results in another gauge transformation of parameter $\bar{\bar{\Lambda}}^\pm$,
\begin{align}
	\label{eq: Conmutador de variaciones 1}
	\comm{\delta}{\bar{\delta}}a^\pm=\bar{\bar{\delta}}a^\pm\,,
\end{align}
where $\bar{\bar{\Lambda}}$ is given by
\begin{align}
	\label{eq: Redefinicion xi}
	\bar{\bar{\Lambda}}^\pm=\delta \bar{\Lambda}^\pm-\bar{\delta}\Lambda^\pm+\comm{\Lambda^\pm}{\bar{\Lambda}^\pm}\,.
\end{align}

Now, let us assume that $\Lambda$ and $\bar \Lambda$ belong to the class of transformations generated by \eqref{las cargas}. Each transformation \eqref{deltas} is canonically generated by $\delta a = \{Q[\Lambda],a\}$ and $\bar \delta a = \{Q[\bar \Lambda],a\}$, where $Q[\Lambda]$ and $Q[\bar \Lambda]$ correspond to some linear combinations of \eqref{las cargas}. It follows from the Jacobi identity that the commutator $\comm{\delta}{\bar{\delta}}a^\pm$ is generated by 
\begin{align}
	\comm{\delta}{\bar{\delta}}a^\pm=\left\{\left\{Q^\pm[\Lambda],{Q[\bar \Lambda]}^\pm\right\},a^\pm\right\}\,.
\end{align}
As the generators satisfy the involution condition \eqref{eq: Asymptotic algebra}, the right-hand side of the latter equation vanishes, indicating that $\bar{\bar \delta} a=0$, or equivalently
\begin{align}
	d\bar{\bar{\Lambda}}^\pm+\comm{a^\pm}{\bar{\bar{\Lambda}}^\pm}=0\,.
\end{align}
Since this relation holds for arbitrary connections $a^\pm$, we conclude that
\begin{align}\label{identity}
	\delta\bar{\Lambda}^\pm-\bar{\delta}\Lambda^\pm+\comm{\Lambda^\pm}{\bar{\Lambda}^\pm}=0\,.
\end{align}
In summary, considering the transformations \eqref{deltas}, the preceding identity remains valid for any pair of gauge transformations whose generators Poisson-commute. Since the temporal component of the connection, $a_t$, belongs to the class of allowable gauge parameters \eqref{at}, the above identity encompasses \eqref{eq: Gauge transformation t} as a specific case, as anticipated.

\subsection{Stationary black hole solution and its thermodynamics}\label{Subsection: BTZ black hole with Dym's boundary conditions and its thermodynamics}

The third requirement for a well-defined boundary condition is to encompass physically sensible gravitational configurations, such as black holes.

Solutions to the equations of the Dym hierarchy \eqref{eq: Dym hierarchy} have been extensively explored in the literature~\cite{Fuchssteiner_1992,Mokhtari2011,SolitonsHD, Novikov1999,RoyChowdhuryMukherjee1984,Bordag:1995dp,alezGaxiola2018ITERATIVEMF,XIAO2019123248,Assabaai2022ExactSO,LI2020106276,Li2017,DMITRIEVA199365,shunmugarajan2016efficient}. A prerequisite for these solutions to represent stationary black holes is that they must be independent of time and exhibit periodicity in the angular coordinate. For instance, solitons typically do not meet such criteria. Therefore, our focus lies in the periodic solutions to the equation
\begin{equation}
\label{estacionaria}
  0 = \d_\phi \cL^\pm -  \d^3_\phi \left( \frac{\delta H_N^\pm}{\delta \cL^\pm}\right)\,.
\end{equation}
To solve the latter, we need to delve into the toolkit provided by the literature on integrable systems, work which will be pursued elsewhere. For our current objectives, we will concentrate on a much simpler family of solutions, given by constant functions
\begin{align}\label{constant}
	\LLpm(t,\phi)=\cL^\pm\,.
\end{align}
Note that this solution satisfies the stationary equation for any choice of $N$. These solutions correspond to black hole configurations, as shown in this section using the metric formulation.

Before proceeding, a few comments are in order. Firstly, note that $\mathcal{L}^\pm=0$ is a singularity of the dynamical equation for any $N \ge 1$ (see \ref{Appendix: Subsection: List}); hence, it is not encompassed in the set of allowed solutions. As shown below, this implies that extremal black holes are not part of this class of solutions. Secondly, due to the presence of square roots in the functions $\mu^\pm$, negative values for $\mathcal{L}^\pm$ must be avoided to prevent complex configurations. As a result, global AdS$_3$ spacetime is not a member of this family of solutions. Finally, it is worth noting that in this class of solutions, the only non-vanishing conserved quantities are $H_0^\pm$ and $H_1^\pm$. As a result of this, the functions $\mu^\pm$ become constants
\begin{align}\label{muconstante}
	\mu^\pm=1+\frac{1}{2 c^N \sqrt{\cL^\pm}}\, .
\end{align}

We initiate the analysis of the geometry by constructing the metric, in accordance to the relation
\begin{align}
	g_{\mu\nu}=\frac{\ell^2}{2}\mean{\left(A_\mu^+-A_\mu^-\right)\left(A_\nu^+-A_\nu^-\right)}\,,
\end{align}
where $A^\pm=b^{-1}_\pm\left(d+a^\pm\right)b_\pm$ and the radial group element is chosen as $b_\pm(\rho)=\exp\left[\pm\log\left(\frac{\rho}{\ell}\right)L_0\right]$. Thus, the stationary line element reads
\begin{align}
	\label{eq: Dym black hole stationary}
	\begin{split}
		ds^2&=-\left(\frac{\ell}{2c\rho}\mu^-\LL^--\frac{\rho}{\ell}\mu^+\right)\left(\frac{\ell}{2c\rho}\mu^+\LL^+-\frac{\rho}{\ell}\mu^-\right)dt^2+\frac{\ell^2 d\rho^2}{\rho^2}\\
		&\hspace{1cm}+\left\{\bigg[\frac{\rho^2}{\ell}+\frac{\ell}{c}\left(1+\frac{\ell^2}{4c\rho^2}\LL^-\right)\LL^+\bigg]\mu^+-\bigg[\frac{\rho^2}{\ell}+\frac{\ell}{c}\left(1+\frac{\ell^2}{4c\rho^2}\LL^+\right)\LL^-\bigg]\mu^-\right\}dt d\phi\\
		&\hspace{7cm}+\left(\rho+\frac{\ell^2}{2c\rho}\LL^+\right)\left(\rho+\frac{\ell^2}{2c\rho}\LL^-\right)d\phi^2\,.
	\end{split}
\end{align}

It is convenient to rewrite the latter metric using Schwarzschild-like coordinates. For this purpose, it is helpful to redefine the radial coordinate according to
\begin{align}
	r^2=\left(\rho+\frac{\ell^2}{2c\rho}\LL^+\right)\left(\rho+\frac{\ell^2}{2c\rho}\LL^-\right)\,,
\end{align}
through which it can be expressed as
\begin{align}
	\label{eq: ADM metric}
	ds^2= -N^2(r) dt^2 + \frac{dr^2}{f^2(r)} + r^2\left(N^\phi(r)dt + d\phi\right)^2\,.
\end{align}
The lapse and shift functions $N(r)$ and $N^\phi(r)$ are given by
\begin{align}
	N(r)=\left(\frac{\mup+\mum}{2}\right) f(r)\,,\quad 	N^\phi(r)=\frac{\mup-\mum}{2\ell}+ \frac{\ell\left(\mup+\mum\right)\left(\LLp-\LLm\right)}{4c r^2}\, ,
\end{align}
while the function $f(r)$ corresponds to
\begin{align}
  f^2(r)=\frac{r^2}{\ell^2}-\frac{\LLp+\LLm}{c} + \frac{\ell^2\left(\LLp-\LLm\right)^2}{4c^2r^2}\,.
\end{align}
The horizon is determined by the condition $N(r)=0$, indicating the positions of both the inner and outer horizons
\begin{align}
	r_\pm^2=\frac{\ell^2}{2c}\left(\cL^+ + \cL^-\pm 2 \sqrt{\LLp \LLm}\right)\,.
\end{align}
Extremality occurs when either $\cL^+=0$ or $\cL^-=0$, a condition not supported in the solution spectrum. Consequently, this family of solutions does not include extremal black holes.

Before adressing the black hole conserved charges and thermodynamics, it's important to note that global AdS$_3$ cannot be encompassed within this set of solutions due to the square roots in $\mu^\pm$, which prohibit negative values for $\cL^\pm$. This is in line with the boundary conditions, which permit the geometry to fluctuate at the leading order near the boundary. As seen from \eqref{eq: ADM metric}, the spatial part of the metric takes the form $g_{ij}=g_{ij}^{AdS}+...$, where the ellipsis denotes subleading terms near the boundary, a consequence of the choice \eqref{bcspatial} for the gauge fields. However, the temporal components fluctuate at the leading order due to the choice of the Lagrange multiplier \eqref{at}. Consequently, it is not expected for the three-dimensional AdS spacetime to be part of this family of solutions.

\begin{figure}
  \begin{center}
    \begin{tikzpicture}
      \def\a{3} 
      \def\b{0.5}
      \draw[pattern=north east lines,line width=1.2, dashed, name path=curve1] plot[variable=\t, domain=3:0, samples=100] ({-\t},{\b*\t+\b*\a*sqrt(\t)}) -- plot[variable=\t, domain=0:3, samples=100] ({\t},{\b*\t+\b*\a*sqrt(\t)});
      \draw[latex-latex] (-3,0) -- (3,0);
      \draw[-latex,name path=curve3] (0,0) -- (0,5);
      
      \node at (-0.3,5) {$M$};
      \node at (3.2,0) {$J$};
      \node at (3.2,4.4) {\footnotesize $\cL^-=0$};
      \node at (-3.2,4.4) {\footnotesize $\cL^+=0$};
    \end{tikzpicture}
  \end{center}
  \caption{Black hole solution spectrum in terms of the mass $M$ and angular momentum $J$. The hatched area corresponds to black holes, while the dashed lines represent extremal configurations, which are not permissible within this solution family.}
  \label{phasespace}
\end{figure}
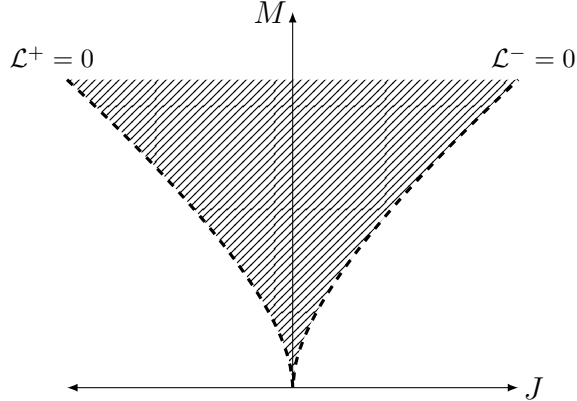

The black hole's mass and angular momentum are conserved quantities associated with the Killing vectors $\partial_t$ and $-\partial_\phi$, respectively. In the Chern-Simons formulation, these are expressed as
\begin{align}
\delta Q^\pm\left[\xi\right]=\frac{k}{2\pi}\oint_{\rho\to\infty} d\phi\,{\xi}^\mu\bigg\langle a^\pm_\mu\delta a^\pm_\phi\bigg\rangle\,,
\end{align}
where $\xi$ represents the Killing vector. Considering both $\pm$ sectors, the total charge is $Q=Q^+-Q^-$. Consequently, the mass and angular momentum correspond to
\begin{align}
\label{eq: M and J}
M=\frac{2\pi}{\ell}\left[\left(\LL^++\LL^-\right)+\frac{1}{c^N}\left(\sqrt{\LL^+}+\sqrt{\LL^-}\right)\right]\,,\quad J=-2\pi\left(\LL^+-\LL^-\right)\,.
\end{align}
Figure \ref{phasespace} illustrates the black hole spectrum in terms of $M$ and $J$.

The thermal properties of the black hole can be examined within the Euclidean counterpart of the metric \eqref{eq: ADM metric}. The Hawking temperature $\beta$ and horizon angular velocity $\Omega_h$ are selected to regularize the Euclidean version of metric \eqref{eq: ADM metric} in the near-horizon sector (see appendix \ref{Appendix: Euclidean analysis} for details). This yields
\begin{subequations}
  \begin{align}\label{eq: Beta}
    \beta&=\frac{4\pi\ell^2 r_+}{\left(\mup+\mum\right)\left(r_+^2-r_-^2\right)}=\frac{\sqrt{2c}\ell \pi(\sqrt{\cL^+} + \sqrt{\cL^-})}{\left(\mup+\mum\right)\sqrt{\cL^-}\sqrt{\cL^+}} \,,\\\label{eq: Omega}
    \Omega_h&=-N^\phi(r_+)=\frac{\mum \sqrt{\cL^-}-\mup \sqrt{\cL^+}}{ \ell (\sqrt{\cL^+} + \sqrt{\cL^-})}\,.
  \end{align}
\end{subequations}

The black hole entropy can be deduced from the boundary term $B$ needed to regularize the Euclidean action. Through a straightforward calculation detailed in Appendix \ref{Appendix: Euclidean analysis}, the variation $\delta B$ takes the following form
\begin{align}
\label{eq: Variation B GR}
\delta B=i\beta\left(\delta M-\frac{1}{\beta}\delta S-\Omega_h\delta J\right)\,,
\end{align}
where the entropy $S$ is expressed as
\begin{align}
S=\frac{2 \pi r_+}{4G}=\frac{\pi\ell\left(\sqrt{\LLp}+\sqrt{\LLm}\right)}{2G\sqrt{2c}}\,,
\end{align}
as expected by the universality of the Bekenstein-Hawking result.

\subsection{More general solutions}

Having introduced the physically sensible black hole solution of the previous section, it is worth exploring more general solutions. Although a difficult task due to the nonlinearity of the equations of motion, we briefly mention some avenues worth exploring.

A simple but enlightening route can be pursued by considering linearized solutions around the black hole background \eqref{eq: Dym black hole stationary}. Consider functions $\cL^\pm(t,\phi)$ of the form
\begin{equation}
  \cL^\pm(t,\phi)=\cL^\pm + h^\pm(t,\phi),
\end{equation}
where $\cL^\pm$ are constants as in \eqref{constant}, while the perturbations $h^\pm$ are small $h^\pm\ll 1$. With this choice, the functions $\mu^\pm(t,\phi)$ are given to first order in $h^\pm$ by
\begin{equation}
  \mu^\pm(t,\phi)=\mu^\pm - \Xi^\pm(t,\phi)\,.
\end{equation}
The constants $\mu^\pm$ are those found in \eqref{muconstante}, while the small functions $\Xi^\pm$ correspond to
\begin{equation}
  \Xi^\pm=\sum_{n=1}^{N} \frac{\sigma_n^\pm}{c^{N-n+1}}\partial_\phi^{2n-2}h^\pm\,,
\end{equation}
where $\sigma^\pm_n\equiv  \left(2^{n+1}{\cL^\pm}^{n+1/2}\right)^{-1}$. This induces a perturbation of the black hole geometry $g_{\mu\nu}^{\mathrm{bh}}$ given in \eqref{eq: Dym black hole stationary} of the form $g_{\mu\nu}=g_{\mu\nu}^{\mathrm{bh}}+h_{\mu \nu}$, where the non vanishing coefficients of $h_{\mu\nu}$ are
\begin{align}
  \begin{split}
  	\begin{split}
  	h_{tt}=&   \left(   \frac{\rho}{\ell}\mu^- -\frac{\ell}{2c\rho}\cL^+ \mu^+   \right)\left[ \frac{\rho}{\ell}\Xi^+ + \frac{\ell}{2c\rho}  \left( c \partial_\phi^2 \Xi^-  -\cL^- \Xi^-+h^- \mu^-       \right)        \right]     \\
  +&\left(  \frac{\rho}{\ell}\mu^+  -\frac{\ell}{2c\rho}\cL^- \mu^-  \right)\left[\frac{\rho}{\ell}\Xi^- + \frac{\ell}{2c\rho}  \left( c \partial_\phi^2 \Xi^+  -\cL^+ \Xi^++h^+ \mu^+       \right)  \right]\,,   
  \end{split}\\
  h_{t\rho}=&\frac{\ell}{2 \rho}\left( \partial_\phi \Xi^+ - \partial_\phi \Xi^-\right)\,, \\
	\begin{split}
	h_{t\phi}=&  \frac{\ell^2}{4c\rho}  \left[  \left(  \frac{\ell}{2c\rho}\cL^+ \mu^+-\frac{\rho}{\ell}\mu^-     \right) h^- -    \left(  \frac{\ell}{2c\rho}\cL^- \mu^--\frac{\rho}{\ell}\mu^+     \right) h^+ \right] \\
&+\left( \frac{\ell^2}{4c\rho}\cL^-  + \frac{\rho}{2}    \right)\left[ \frac{\rho}{\ell}\Xi^- + \frac{\ell}{2c\rho}  \left( c \partial_\phi^2 \Xi^+  -\cL^+ \Xi^++h^+ \mu^+       \right)        \right]  \\
       & -\left( \frac{\ell^2}{4c\rho}\cL^+  + \frac{\rho}{2}    \right)\left[ \frac{\rho}{\ell}\Xi^+ + \frac{\ell}{2c\rho}  \left( c \partial_\phi^2 \Xi^-  -\cL^- \Xi^-+h^- \mu^-       \right)  \right]\,,  
	\end{split}\\
h_{\phi\phi}=&\frac{\ell^2}{2 c} \left[ \left( 1+\frac{\ell^2}{2 c \rho^2}\cL^- \right) h^++\left( 1+\frac{\ell^2}{2 c \rho^2}\cL^+ \right) h^- \right]\,.
  \end{split}
\end{align}

The equations of motion, i.e., the $N$-th member of the Dym hierarchy \eqref{dymhierarchy}, reduce to the linear equations
\begin{equation}\label{eqh}
  \pm\ell\partial_t h^\pm =\partial_\phi h^\pm+\sigma_N^\pm \partial_\phi^{2N+1} h^\pm\,.
\end{equation}
These correspond to combinations of anisotropic chiral bosons, studied in \cite{Fuentealba:2019oty}. The general solution to \eqref{eqh} with periodic boundary conditions is a linear combination of chiral movers
\begin{equation}\label{fmodes}
   h_k^\pm = \exp\left[i\left(k \phi + \omega_k^\pm t   \right) \right]\,, \quad k \in \mathbb{Z}\,,
 \end{equation}
obeying the dispersion relation
 \begin{equation}
   \pm \ell \omega_k^\pm = k + (-1)^N \sigma^\pm k^{2N+1}\,.
 \end{equation}
 The phase velocity is given by $v_p^\pm=\pm \frac{1}{\ell}\left[1+ (-1)^N \sigma^\pm k^{2N}     \right]$. Since $2N$ is an even integer, all Fourier modes \eqref{fmodes} propagate in the same direction, while the opposite chiralities $\pm$ do it in opposite directions. Note that the magnitude of the phase velocity depends on the mass and angular momentum of the background black hole through the constant $\sigma_N^\pm$. Although the presented solution is valid only at the linear level, the dispersion relation does codify the nonlinearities of the Dym equation. These perturbations are reminiscent of the Bañados geometries~\cite{Banados:1998gg} in the case of Brown-Henneaux boundary conditions. These have been studied in the context of holography (see, for example, \cite{OColgain:2016msw,Sheikh-Jabbari:2016znt,Roberts:2012aq}). The anisotropic chiral bosons have also been used to calculate the transport properties of three-dimensional black holes and their connection to memory effects~\cite{Cardenas:2021sun}. It is also worth mentioning that these can be consistently quantized, showing interesting connections to results in number theory~\cite{Fuentealba:2019oty}.

Finding an explicit, fully nonlinear solution is a difficult task, even for the simplest equation in the hierarchy \eqref{eq: Dym equation}, corresponding to $N=1$. Several efforts can be found in the literature \cite{Fuchssteiner_1992,Mokhtari2011,SolitonsHD, Novikov1999,RoyChowdhuryMukherjee1984,Bordag:1995dp,alezGaxiola2018ITERATIVEMF,XIAO2019123248,Assabaai2022ExactSO,LI2020106276,Li2017,DMITRIEVA199365,shunmugarajan2016efficient}, and they mostly lead to implicit solutions. A rather general, although implicit, solution can be found by direct integration. First, consider the field redefinition
\begin{equation}
  \cL^\pm(t,\phi)=\frac{1}{2^{4/3}{u^\pm}^2(t,\phi)},
\end{equation}
which transforms the Dym equation \eqref{eq: Dym equation} to the form
\begin{equation}
  \pm\ell\partial_t u^\pm=\partial_\phi u^\pm +{u^\pm}^3 \partial_\phi^3u^\pm .
\end{equation}
We focus on the family of traveling wave solutions $u^\pm(t,\phi)=u^\pm(\chi^\pm)$, where $\chi^\pm\equiv \phi \pm v \frac{t}{\ell}$ and $v$ is a constant. The equation further reduces to
\begin{equation}
  0= (1- v){u^\pm}' +{u^\pm}^3 {u^\pm}''',
\end{equation}
where the primes represent the derivative with respect to $\chi^\pm$. It can be integrated three times following the procedure described in \cite{Mokhtari2011}. The solution is given in the implicit form
\begin{equation}
   \left[  F\left(  \operatorname{arcsin}\left (\sqrt{\frac{4A}{\gamma_-}}\sqrt{u^\pm}\right)
     ,\frac{\gamma_-}{\gamma_+}\right)-E\left( \operatorname{arcsin}\left(\sqrt{\frac{4A}{\gamma_-}}\sqrt{u^\pm}\right)
 ,\frac{\gamma_-}{\gamma_+}\right)
\right]^2
=\frac{2 A^2 }{\gamma_+}(\chi^\pm+ C)^2,
\end{equation}
where $\gamma_{\pm}=-B \pm \sqrt{8(1-v)A+B^2}$,  while $A$, $B$ and $C$ are arbitrary integration constants. The functions $F$ and $E$ correspond to incomplete elliptic integrals of the first and second kind, respectively. Several particular cases can be found on \cite{Mokhtari2011}, along with other methods of integration. By setting $v=0$, the above expression is also a solution to the stationary equation \eqref{estacionaria}, for $N=1$. Periodic solutions are also included in the previous expression \cite{Bordag:1995dp,DMITRIEVA199365}. It will be explored somewhere else if there is a region in parameter space on which such solutions represent black holes.

\section{Final comments}

We have shown the construction of Dym boundary conditions for AdS$_3$ pure gravity and established their well-defined nature. Notably, these conditions give rise to an infinite and abelian asymptotic symmetry group, where the generators correspond to the Dym conserved charges, and encompass black hole solutions within their spectrum.

We conclude this work with a discussion of avenues worth exploring.

It is important to note that the choice of the integration constants $\alpha_n^\pm$ is arbitrary and was inspired by the goal of constructing deformations of Brown-Henneaux boundary conditions. However, as mentioned earlier, alternative possibilities exist. In particular, selecting $\alpha_0^\pm=0$ leads to the dynamical equation
\begin{equation}
\pm \ell\d_t\cL^\pm = - \d^3_\phi \left( \frac{\delta H_N^\pm}{\delta \cL^\pm}\right)\,,
\end{equation}
which enhances symmetry to include Lifshitz scaling. This might allow to explore Lifshitz holography within this context~\cite{Taylor:2015glc,Gonzalez:2011nz,Hartong:2014oma,Figueroa-OFarrill:2022kcd}.

It is worth mentioning that the whole solution spectrum of the Dym boundary conditions cannot be reached by applying the allowed gauge transformations \eqref{transformations} to the zero-mode solutions studied in section \ref{Subsection: BTZ black hole with Dym's boundary conditions and its thermodynamics}, unlike the case of Brown-Henneaux boundary conditions. In the latter, solutions within the spectrum manifest as chiral modes oscillating across an identified global AdS$_3$ spacetime. Such structure is rooted in the asymptotic symmetries, namely the 2D conformal group, and the linearity of the boundary dynamics, governed by the chiral boson equation. However, these considerations manifest differently in the Dym case. Due to the involution of $H_n$ \eqref{eq: Asymptotic algebra}, the set of conserved charges of two solutions related by a large gauge transformation must coincide. Hence, black holes with nontrivial Dym charges lie on a different orbit than the zero-mode solution and must be found by solving the differential equation \eqref{eq: Dym hierarchy}, which is challenging due to the equation's nonlinearity. A quick review of the literature reveals that finding solutions to the Dym equation (or any integrable system) requires powerful methods such as Darboux and Backlund transformations, consistent Tanh expansion, among others, often yielding particular cases that obscure the systematic construction of the solution spectrum \cite{Fuchssteiner_1992,Mokhtari2011,SolitonsHD, Novikov1999,RoyChowdhuryMukherjee1984,Bordag:1995dp,alezGaxiola2018ITERATIVEMF,XIAO2019123248,Assabaai2022ExactSO,LI2020106276,Li2017,DMITRIEVA199365,shunmugarajan2016efficient}. Therefore, in this work, we focus solely on studying the properties of the zero-mode solution, as presented in section \ref{Subsection: BTZ black hole with Dym's boundary conditions and its thermodynamics}, while the structure of the solution spectrum and its relation to large gauge transformations will be addressed elsewhere.

The holographic implications of this family of boundary conditions will be further explored in subsequent work. The recursive procedure used to construct the Dym boundary conditions is exact and valid for any value of the central charge, allowing for exploration of limiting cases. By noting that the expansion in \eqref{expansion} involves negative integer powers of $c$, it is tempting to speculate that the holographic properties are somehow linked to a boundary conformal theory at large central charge, which has been extensively studied in the context of holography~\cite{El-Showk:2011yvt,Hartman:2013mia,Fitzpatrick:2016thx,Malvimat:2017yaj,Dymarsky:2018iwx,Brehm:2019fyy,Chandra:2022bqq,Benjamin:2023uib}. Such a connection will be pursued elsewhere.

Finally, an extension of this work can be pursued for higher-order algebras, such as $sl(N,R)$, potentially establishing connections between higher spin gravity and a broader hierarchy of integrable equations that includes the Dym hierarchy as a particular case.

\paragraph{Acknowledgments} 

The authors express their gratitude to Marcela Cárdenas, Crist\'obal Corral, Francisco Correa, José Figueroa, Hernán A. González, Luis Guajardo, and Ricardo Troncoso for their insightful comments. This work has received support from FONDECYT grants 1231810, 3230618, ANID Beca Doctorado Nacional 21232318, and DICYT 042231PR PS 530. K.L. would like to extend thanks to the Theoretical and Mathematical Physics group at Université libre de Bruxelles for their warm hospitality, where part of this work was carried out. Furthermore, M.P. and F.R. also acknowledge Centro de Estudios Científicos (CECs) at Universidad San Sebastián for their hospitality during part of this work. 

\appendix
\section{Chern-Simons formulation of AdS$_3$ gravity}\label{Appendix: Section: Chern-Simons Gravity}
We provide a concise overview of the Chern-Simons formulation of AdS$_3$ gravity, focusing solely on essential details to ensure the self-contained nature of the preceding presentation.

$(2+1)$ dimensional General Relativity with a negative cosmological constant can be written as the difference of two Chern-Simons action~\cite{Achucarro:1986uwr,Witten:1988hc}
\begin{align}
	\label{eq: Hamiltonian action}
	I=I_H[A^+]-I_H[A^-]\,,\quad 
	I_H[A]=-\frac{k}{4\pi}\int dt\,d^2x\,\epsilon^{ij}\mean{A_i \dot{A}_j-A_t F_{ij}}+ B\,,
\end{align}
where $F_{ij}^\pm=\partial_i A^\pm_j-\partial_j A_i^\pm+\comm{A_i^\pm}{A_j^\pm}$. The Chern-Simons level is given by  $k=\ell/4G$, where $\ell$ stands for the AdS$_3$ radius while $G$ is the three-dimensional Newton constant. The fields $A^\pm$ are $sl(2,\mathbb{R})$ connections. The three algebra generators $L_n$, where $n\in\left\{0,\pm 1\right\}$, fulfill the algebra
\begin{align}
	\comm{L_m}{L_n}=(m-n)L_{m+n}\,.
\end{align}
The nonvanishing components of the bilinear product $\mean{\,,}$ are
\begin{align}
	\mean{L_0,L_0}=\frac{1}{2}\,,\quad \mean{L_1,L_{-1}}=-1\,.
\end{align}
The boundary term $B$ in (\ref{eq: Hamiltonian action}) is constructed to assure the differentiability of the action, yielding
\begin{align}
	\label{eq: Boundary term}
	\delta B=-\frac{k}{2\pi}\int_{\rho\to\infty} dt\,d\phi\,\mean{A_t\delta A_\phi}\,.
\end{align}
Therefore, adequate boundary conditions must be inforced in order for $\delta B$ to be integrable. 

The infinitesimal gauge transformation
\begin{align}
	\delta A=d\Lambda+\comm{A}{\Lambda}\,,
\end{align}
is generated by the smeared integral
\begin{align}
	G\left[\Lambda\right]=\frac{k}{4\pi}\int d^2x\,\epsilon^{ij}\mean{\Lambda F_{ij}}+Q\,,
\end{align}
where $ Q$ is a surface term assuring the differentiability of the generator, implying
\begin{align}
	\delta Q\left[\Lambda\right]=-\frac{k}{2\pi}\int_{\rho\to\infty} d\phi\,\mean{\Lambda\delta A_\phi}\,.
\end{align} 
If the gauge parameter $\Lambda$ decay fast enough towards the boundary such that $Q$ vanishes, then the transformation is a \textit{proper gauge transform}. On the other hand, if the boundary behavior of the gauge parameter is such that $Q$ is finite, then the transformation is a \textit{large gauge transformation}~\cite{Benguria:1976in}. More details and discussion on this subject can be found in the review \cite{Banados:2016zim} and references therein.

Finally, the link between Chern-Simons and the metric fields is established through $A^\pm=\omega\pm\frac{e}{\ell}$. Consequently, the metric can be constructed as
\begin{align}
	g_{\mu\nu}=\frac{\ell^2}{2}\mean{\left(A_\mu^+-A_\mu^-\right),\left(A_\nu^+-A_\nu^-\right)}\,.
\end{align}

\newpage
\section{List of Dym conserved charges $\HHpm_n$} \label{Appendix: Subsection: List}

Following the recursive procedure outlined in Section \ref{construction}, we present the explicit form of the first five functionals $\HHpm_n$ along with their respective functional derivatives. In what follows, the prime denotes derivatives with respect to $\phi$. Introducing the notation $\HHpm_n=\int d\phi\,\cH^\pm_n$, the first five charge densities are as follows:
{\begin{align}
		\label{ap: eq: Hamiltonian densities}
		\begin{split}
			\cH^\pm_1&=\sqrt{\LLpm}\,,\\
			\cH^\pm_2&=-\frac{1}{2\sqrt{\LLpm}}\left(\frac{5}{8}\frac{{\LLpm'}^2}{\LLpm^2}-\frac12 \frac{\LLpm''}{\LLpm}\right)\,,\\
			\cH^\pm_3 &= -\frac{1}{2\sqrt{\LLpm}}\left(\frac{385}{256}\frac{{\LLpm'}^4}{\LLpm^5}-\frac{77}{32}\frac{{\LLpm'}^2\LLpm''}{\LLpm^4}+\frac{7}{16}\frac{{\LLpm''}^2}{\LLpm^3}+\frac{7}{12}\frac{\LLpm'\LLpm'''}{\LLpm^3}-\dfrac{1}{12}\frac{\LLpm^{(4)}}{\LLpm^2}\right)\,,\\
			\cH^\pm_4&=-\frac{1}{2\sqrt{\LLpm}}\left(\frac{85085}{4096}\frac{{\LLpm'}^6}{\LLpm^8}-\frac{51051}{1024}\frac{{\LLpm'}^4\LLpm''}{\LLpm^7}+\frac{35607}{1280}\frac{{\LLpm'}^2{\LLpm''}^2}{\LLpm^6}-\frac{671}{320}\frac{{\LLpm''}^3}{\LLpm^5}\right.\\
			&\hspace{2cm}+\frac{429}{32}\frac{{\LLpm'}^3\LLpm'''}{\LLpm^6} -\frac{1419}{160}\frac{\LLpm'\LLpm''\LLpm'''}{\LLpm^5}+\frac{69}{160}\frac{{\LLpm'''}^2}{\LLpm^4}-\frac{165}{64}\frac{{\LLpm'}^2\LLpm^{(4)}}{\LLpm^5}\\
			&\hspace{6cm}\left. +\frac{57}{80}\frac{\LLpm''\LLpm^{(4)}}{\LLpm^4}+\frac{27}{80}\frac{\LLpm'\LLpm^{(5)}}{\LLpm^4}-\frac{1}{40}\frac{\LLpm^{(6)}}{\LLpm^3}\right)\,,\\
			\cH^\pm_5&=-\frac{1}{2\sqrt{\mathcal{L}^\pm}}\left(\frac{185910725}{262144}\frac{ {{\mathcal{L}^\pm}'}^8}{ {\mathcal{L}^\pm}^{11}}-\frac{37182145 }{16384}\frac{{{\mathcal{L}^\pm}''} {{\mathcal{L}^\pm}'}^6}{ {\mathcal{L}^\pm}^{10}}+\frac{9006855}{14336}\frac{ {{\mathcal{L}^\pm}'''} {{\mathcal{L}^\pm}'}^5}{ {\mathcal{L}^\pm}^9}\right.\\
			&\hspace{1cm}+\frac{123740331 }{57344}\frac{{{\mathcal{L}^\pm}''}^2 {{\mathcal{L}^\pm}'}^4}{ {\mathcal{L}^\pm}^9}-\frac{1859715}{14336}\frac{ {{\mathcal{L}^\pm}^{(4)}} {{\mathcal{L}^\pm}'}^4}{ {\mathcal{L}^\pm}^8}+\frac{36465}{1792}\frac{ {{\mathcal{L}^\pm}^{(5)}} {{\mathcal{L}^\pm}'}^3}{ {\mathcal{L}^\pm}^7}\\
			&\hspace{0.5cm}-\frac{3201627}{3584}\frac{ {{\mathcal{L}^\pm}''} {{\mathcal{L}^\pm}'''} {{\mathcal{L}^\pm}'}^3}{ {\mathcal{L}^\pm}^8}+\frac{282711}{3584}\frac{ {{\mathcal{L}^\pm}'''}^2 {{\mathcal{L}^\pm}'}^2}{ {\mathcal{L}^\pm}^7}+\frac{226083}{1792}\frac{ {{\mathcal{L}^\pm}''} {{\mathcal{L}^\pm}^{(4)}} {{\mathcal{L}^\pm}'}^2}{ {\mathcal{L}^\pm}^7}\\
			&-\frac{2145 }{896}\frac{{{\mathcal{L}^\pm}^{(6)}} {{\mathcal{L}^\pm}'}^2}{ {\mathcal{L}^\pm}^6}-\frac{629629 }{1024}\frac{{{\mathcal{L}^\pm}''}^3 {{\mathcal{L}^\pm}'}^2}{ {\mathcal{L}^\pm}^8}+\frac{97383 }{448}\frac{{{\mathcal{L}^\pm}''}^2 {{\mathcal{L}^\pm}'''} {{\mathcal{L}^\pm}'}}{ {\mathcal{L}^\pm}^7}+\frac{11}{56}\frac{{{\mathcal{L}^\pm}^{(7)}} {{\mathcal{L}^\pm}'}}{ {\mathcal{L}^\pm}^5}\\
			&-\frac{1287}{112}\frac{ {{\mathcal{L}^\pm}''} {{\mathcal{L}^\pm}^{(5)}} {{\mathcal{L}^\pm}'}}{ {\mathcal{L}^\pm}^6}-\frac{3861}{224}\frac{ {{\mathcal{L}^\pm}'''} {{\mathcal{L}^\pm}^{(4)}} {{\mathcal{L}^\pm}'}}{ {\mathcal{L}^\pm}^6}+\frac{180323 }{7168}\frac{{{\mathcal{L}^\pm}''}^4}{{\mathcal{L}^\pm}^7}+\frac{253}{448}\frac{ {{\mathcal{L}^\pm}^{(4)}}^2}{{\mathcal{L}^\pm}^5}\\
			&\hspace{1cm}+\frac{209}{224}\frac{ {{\mathcal{L}^\pm}'''} {{\mathcal{L}^\pm}^{(5)}}}{{\mathcal{L}^\pm}^5}+\frac{121}{224}\frac{ {{\mathcal{L}^\pm}''} {{\mathcal{L}^\pm}^{(6)}}}{ {\mathcal{L}^\pm}^5}-\frac{1}{112}\frac{{{\mathcal{L}^\pm}^{(8)}}}{ {\mathcal{L}^\pm}^4}-\frac{13299}{896 }\frac{ {{\mathcal{L}^\pm}''} {{\mathcal{L}^\pm}'''}^2}{{\mathcal{L}^\pm}^6}\\
			&\hspace{8cm}\left.-\frac{10725}{896}\frac{ {{\mathcal{L}^\pm}''}^2 {{\mathcal{L}^\pm}^{(4)}}}{ {\mathcal{L}^\pm}^6}\right)\,.
		\end{split}
\end{align}}%

The corresponding functionals derivatives are:
\begin{align}
	\label{ap: eq: Variational H}
\begin{split}
	\frac{\delta H_1^\pm}{\delta \cL^\pm} &= \frac{1}{2\sqrt{\LLpm}}\,,\\
	\frac{\delta H_2^\pm}{\delta \cL^\pm} &= \frac{1}{2\sqrt{\LLpm}}\left(\frac{5}{16}\frac{{\LLpm}'^2}{\LLpm^3}-\frac14 \frac{\LLpm''}{\LLpm^2}\right)\,,\\
	\frac{\delta H_3^\pm}{\delta \cL^\pm} &= \frac{1}{2\sqrt\LLpm}\left( \frac{1155}{512}\frac{{\LLpm'}^4}{ \LLpm^6}-\frac{231}{64}\frac{{\LLpm'}^2\LLpm''}{\LLpm^5}+\frac{21}{32}\frac{{\LLpm''}^2}{\LLpm^4}+\frac{7}{8}\frac{\LLpm'\LLpm'''}{\LLpm^4} - \frac{1}{8}\frac{\LLpm^{(4)}}{\LLpm^3}\right)\,,\\
	\begin{split}
	\frac{\delta H_4^\pm}{\delta \cL^\pm} &=\frac{1}{2\sqrt{\LLpm}}\left(  \frac{425425}{8192}\frac{{\LLpm'}^6}{\LLpm^9}-\frac{255255}{2048}\frac{{\LLpm'}^4\LLpm''}{\LLpm^8}+\frac{35607}{512}\frac{{\LLpm'}^2{\LLpm''}^2}{\LLpm^7}-\frac{671}{128}\frac{{\LLpm''}^3}{\LLpm^6}\right.\\
		&\hspace{1.5cm}+ \frac{2145}{64}\frac{{\LLpm'}^3\LLpm'''}{\LLpm^7}-\frac{1419}{64}\frac{{\LLpm'}{\LLpm''}{\LLpm'''}}{\LLpm^6}+\frac{69}{64}\frac{{\LLpm'''}^2}{\LLpm^5}-\frac{825}{128}\frac{{\LLpm'}^2\LLpm^{(4)}}{\LLpm^6}\\
		&\hspace{5cm}\left.+ \frac{57}{32}\frac{\LLpm''\LLpm^{(4)}}{\LLpm^5}+\frac{27}{32}\frac{\LLpm'\LLpm^{(5)}}{\LLpm^5}-\frac{1}{16}\frac{\LLpm^{(6)}}{\LLpm^4}\right)\,,\\
	\fdv{\HHpm_5}{\LLpm}&=\frac{1}{2 \sqrt{{\mathcal{L}^\pm}}} \left(\frac{1301375075}{524288}\frac{ {{\mathcal{L}^\pm}'}^8}{ {\mathcal{L}^\pm}^{12}}-\frac{260275015}{32768}\frac{ {{\mathcal{L}^\pm}''} {{\mathcal{L}^\pm}'}^6}{ {\mathcal{L}^\pm}^{11}}+\frac{9006855}{4096}\frac{ {{\mathcal{L}^\pm}'''} {{\mathcal{L}^\pm}'}^5}{ {\mathcal{L}^\pm}^{10}}\right.\\
	&\hspace{2cm}+\frac{123740331}{16384}\frac{ {{\mathcal{L}^\pm}''}^2 {{\mathcal{L}^\pm}'}^4}{ {\mathcal{L}^\pm}^{10}}-\frac{1859715}{4096}\frac{ {{\mathcal{L}^\pm}^{(4)}} {{\mathcal{L}^\pm}'}^4}{ {\mathcal{L}^\pm}^9}+\frac{36465}{512}\frac{ {{\mathcal{L}^\pm}^{(5)}} {{\mathcal{L}^\pm}'}^3}{ {\mathcal{L}^\pm}^8}\\
	&\hspace{1cm}-\frac{3201627}{1024}\frac{ {{\mathcal{L}^\pm}''} {{\mathcal{L}^\pm}'''} {{\mathcal{L}^\pm}'}^3}{ {\mathcal{L}^\pm}^9}+\frac{282711}{1024}\frac{ {{\mathcal{L}^\pm}'''}^2 {{\mathcal{L}^\pm}'}^2}{ {\mathcal{L}^\pm}^8}+\frac{226083}{512}\frac{ {{\mathcal{L}^\pm}''} {{\mathcal{L}^\pm}^{(4)}} {{\mathcal{L}^\pm}'}^2}{ {\mathcal{L}^\pm}^8}\\
	&\hspace{2cm}-\frac{2145}{256}\frac{ {{\mathcal{L}^\pm}^{(6)}} {{\mathcal{L}^\pm}'}^2}{ {\mathcal{L}^\pm}^7}-\frac{4407403}{2048}\frac{ {{\mathcal{L}^\pm}''}^3 {{\mathcal{L}^\pm}'}^2}{ {\mathcal{L}^\pm}^9}+\frac{97383 }{128}\frac{{{\mathcal{L}^\pm}''}^2 {{\mathcal{L}^\pm}'''} {{\mathcal{L}^\pm}'}}{ {\mathcal{L}^\pm}^8}\\
	&\hspace{2cm}+\frac{11 }{16}\frac{{{\mathcal{L}^\pm}^{(7)}} {{\mathcal{L}^\pm}'}}{ {\mathcal{L}^\pm}^6}-\frac{1287}{32}\frac{ {{\mathcal{L}^\pm}''} {{\mathcal{L}^\pm}^{(5)}} {{\mathcal{L}^\pm}'}}{ {\mathcal{L}^\pm}^7}-\frac{3861}{64}\frac{ {{\mathcal{L}^\pm}'''} {{\mathcal{L}^\pm}^{(4)}} {{\mathcal{L}^\pm}'}}{ {\mathcal{L}^\pm}^7}\\
	&\hspace{2cm}+\frac{180323}{2048}\frac{ {{\mathcal{L}^\pm}''}^4}{ {\mathcal{L}^\pm}^8}+\frac{253}{128}\frac{ {{\mathcal{L}^\pm}^{(4)}}^2}{ {\mathcal{L}^\pm}^6}+\frac{209}{64}\frac{{{\mathcal{L}^\pm}'''} {{\mathcal{L}^\pm}^{(5)}}}{ {\mathcal{L}^\pm}^6}+\frac{121}{64}\frac{ {{\mathcal{L}^\pm}''} {{\mathcal{L}^\pm}^{(6)}}}{ {\mathcal{L}^\pm}^6}\\
	&\hspace{3.5cm}\left.-\frac{1}{32}\frac{{{\mathcal{L}^\pm}^{(8)}}}{ {\mathcal{L}^\pm}^5}-\frac{13299 }{256 }\frac{{{\mathcal{L}^\pm}''} {{\mathcal{L}^\pm}'''}^2}{{\mathcal{L}^\pm}^7}-\frac{10725}{256}\frac{ {{\mathcal{L}^\pm}''}^2 {{\mathcal{L}^\pm}^{(4)}}}{ {\mathcal{L}^\pm}^7}\right)\,.
	\end{split}
\end{split}
\end{align}
Note that these satisfy the following identity~\cite{tu1989} 
\begin{align}
	\HHpm_n=\frac{2}{3-2n}\int d\phi\,\frac{\delta H_n^\pm}{\delta \cL^\pm}    \LLpm\,.
\end{align}

\section{Euclidean analysis of black hole solution}\label{Appendix: Euclidean analysis}

In this analysis, the Euclidean version of the metric \eqref{eq: ADM metric} is examined to extract the thermodynamic properties of the black hole solution.

Consider first the Wick rotation
\begin{align}
\label{ap: eq: Wick rotation}
t\to i\tau\,.
\end{align}
Upon implementing, the metric takes on the Euclidean form
\begin{align}
\label{ap: eq: Euclidean metric 2}
\euclidean{ds^2}= N^2(r) d\tau^2 + \frac{dr^2}{f^2(r)} + r^2\left(\euclidean{N^\phi}(r)d\tau + d\phi\right)^2\,,
\end{align}
where $\euclidean{N^j}=iN^j$.

To determine the temperature and velocity of the horizon, the Euclidean metric near the surface $r=r_+$ is addressed. It results in
\begin{align}
\euclidean{ds^2}\simeq \frac{1}{4}\left(\frac{\mup+\mum}{2}\right)^2  \left( \left.{f^2(r)}'\right|_{r_+} \right)^2    R^2 d\tau^2+dR^2+r_+^2\left(iN^\phi(r_+)d\tau+d\phi\right)^2\,,
\end{align}
Here, the radial coordinate has been redefined as $R=2\sqrt{\frac{(r-r_+)}{\left. {f^2(r)}'\right|_{r_+}}}$. In order for the latter to represent a cylinder of the form $\euclidean{ds^2}=R^2d\theta^2+dR^2+dz^2$, the following identification of coordinates is performed
\begin{align}
\theta=\left(\frac{\mup+\mum}{4}\right)  \left( \left.{f^2(r)}'\right|_{r_+} \right)    \tau\,,\quad dz=r_+\left(iN^\phi(r_+)d\tau+d\phi\right)\,.
\end{align}
The angular periodicity $\theta\sim\theta+2\pi$ implies that the temporal coordinate is identified as $\tau\sim \tau+\beta$, while the absence of periodicity in $z$ (the transverse section) results in $\phi\sim \phi+i\beta\Omega_h$. Consequently, the inverse of the Hawking temperature $\beta$ and the velocity of the horizon $\Omega_h$ are in agreement with \eqref{eq: Beta} and \eqref{eq: Omega}, respectively.

Now, details about the derivation of ~\eqref{eq: Variation B GR} are presented. To determine the variation of the boundary term $\delta B$, the Hamiltonian formulation is employed. The Euclidean Hamiltonian action of General Relativity is expressed as
\begin{align}
\label{eq: Action}
I_E=I_H+B\,.
\end{align}
Focusing our attention solely on the boundary term, its variation is given by~\cite{Regge:1974zd}
\begin{align}
\label{ap: eq: Delta B}
\delta B=-\frac{1}{16\pi G} \int dS^2_l\, \left[G^{ijkl}\left(N_{|k} \delta \gamma_{ij}-N \delta\gamma_{ij|k}\right)-2N_{\E,k} \delta \pi^{kl}_\E-\left(2\pi^{il}_\E N^j_\E-\pi^{ij}_\E N^l_\E\right)\delta\gamma_{ij}\right]\,.
\end{align}
Here, $\gamma_{ij}$ represents the spatial metric, and $\pi^{ij}$ is its conjugate momentum. Additionally, $N$ and $N_E^i$ denote the lapse and shift vectors, respectively.

Considering the stationary and axisymmetric symmetry of the solution \eqref{eq: ADM metric}, the variation of $B$ simplifies to
\begin{align}
  \delta B&=\frac{i}{16\pi G}\int_0^{\beta}d\tau \int_0^{2\pi}d\phi \left.\left(G^{ijkr}N \delta\gamma_{ij|k} +2N_{\E,j} \delta\pi^{rj}_\E\right)\right\rvert_{r=r_+}^{r=\infty}\,.
\end{align}
Expressed in terms of Dym's fields, the above can be written as
\begin{align}
  \begin{split}
    \label{ap: eq: Boundary term reduced}
    \delta B&= \frac{i}{16\pi G} \beta \left(2\pi\right) \left[\frac{1}{c}\left(\mup\delta\LLp +\mum\delta\LLm \right)-\frac{4\pi}{\beta}\delta r_+ \right.\\
            &\hspace{5cm}\left.-\frac{1}{c} \frac{\left(\mup\sqrt{\LLp}-\mum\sqrt{\LLm}\right)}{\sqrt{\LLp}+\sqrt{\LLm}}\left(\delta\LLp-\delta\LLm\right)\right]\,.
  \end{split}
\end{align}
Recognizing the mass and angular momentum \eqref{eq: M and J}, we arrive at \eqref{eq: Variation B GR}, as promised.

\clearpage
\printbibliography

\end{document}